\documentclass[preprint,aps,pre,nofootinbib,showpacs]{revtex4}   

\usepackage{graphicx} 
\usepackage{color}
\usepackage{amssymb}
\usepackage{amsmath}
\usepackage[normalem]{ulem}
\newcommand{\be}{\begin{equation}}
\newcommand{\ee}{\end{equation}}
\newcommand{\ba}{\begin{align}}
\newcommand{\ea}{\end{align}}

\newcommand{\ct}{\tau_c}

\begin{document}

\title{Efficient harmonic oscillator chain energy harvester driven by colored noise}

\author{M.~Romero-Bastida}
\affiliation{SEPI ESIME-Culhuac\'an, Instituto Polit\'ecnico Nacional, Avenida Santa Ana 1000, Colonia San Francisco Culhuac\'an, Delegaci\'on Coyoacan, Distrito Federal 04430, Mexico}
\email{mromerob@ipn.mx}
\author{Juan M. L\'opez}
\email{lopez@ifca.unican.es}
\affiliation{Instituto de F\'\i sica de Cantabria (IFCA), CSIC-Universidad de
Cantabria, E-39005 Santander, Spain}

\date{\today}

\begin{abstract}
We study the performance of an electromechanical harmonic oscillator chain as an energy harvester to extract power from finite-bandwidth ambient random vibrations, which are modelled by colored noise. The proposed device is numerically simulated and its performance assessed by means of the net electrical power generated and its efficiency in converting the external noise-supplied power into electrical power. Our main result is a much enhanced performance, both in the net electrical power delivered and in efficiency, of the harmonic chain with respect to the popular single oscillator resonator. Our numerical findings are explained by means of an analytical approximation, in excellent agreement with numerics. 

\vskip0.3cm

\end{abstract}


\maketitle

\section{Introduction\label{sec:Intro}}

A huge development in the miniaturization capability of electronic devices has been observed in the last few years. However, the energy density available in batteries aimed at providing the powering for such devices has not reached the same rate of improvement when operated in stand-alone configurations~\cite{Paradiso05}. Among various possibilities to solve this, as well as other energy-management related issues, it has been proposed the harvesting of ambient micro-kinetic energy from the environment, mostly available in the form of random vibrations. In fact, a significant amount of kinetic energy is actually present as mechanical displacements characterized by periodic and stochastic components. Additionally, shrinking the dimension of mechanical elements down to the nano-scale results in an increment of the harvesting efficiency in terms of power density and in a significant reduction in mass fabrication costs. Kinetic energy harvesting requires a mechanical system that couples environmental displacements to a transduction mechanism for vibrational to electrical energy conversion. To date, various energy harvesters have been developed that rely on capacitive~\cite{Okamoto12}, inductive~\cite{Kwon13}, and piezoelectric transduction mechanisms~\cite{Anton07,Renno09,Kim11,Zeng13}.

Regardless of the employed transduction mechanism, most of the vibrational energy harvesters ---also known as vibration power generators ---consider a linear spring or single harmonic oscillator as the mechanical element of the device and treat the external vibrations as sinusoidal vibrations. Thus the maximum power is generated when the resonant frequency of the generator matches the ambient vibration frequency, known as resonant energy harvesting~\cite{Beeby06}. Nearly all current vibration transducers operate in this regime~\cite{Spreemann12}. However, this approach presents numerous drawbacks, being one of the most important ones that the linear harvester resonant peak is necessarily very narrow~\cite{Burrow08}, which limits their application in real-world environments with stochastic fluctuations and a continuous spectrum of vibration frequencies~\cite{Wang08a}.

To overcome these difficulties, a different approach based on the exploitation of the properties of non-resonant oscillators, {\it {\it i.e.}} characterized by a non-linear dynamical response, has been proposed~\cite{Cottone09,Gammaitoni09,Erturk09,Harne13,Hosseinloo16,Hosseinloo15}. The main rationale behind this approach is to try to take advantage of the broad bandwidth frequency response associated with nonlinear systems as opposed to the resonant, narrow bandwidth, single-frequency response that characterizes purely harmonic oscillators. If the broadband ambient vibrations are modelled by Gaussian white noise, many important results have been obtained. For example, it has been shown that, if we consider bistable oscillators under proper operating conditions, they can provide better performances compared to those of a linear oscillator in terms of energy extracted from a generic wide spectrum vibration~\cite{Cottone09}. It has also been established, using the Fokker-Planck equation to describe Duffing-type energy harvesters, that the mean power output of the device is not affected by the nonlinearity of the spring~\cite{Daqaq10,Green12}. Also, the upper bound on the power output of generic nonlinear energy harvesters driven by Gaussian white noise has been obtained and it has been shown that, subject to mild restrictions on the device parameters, it is always possible to find an {\em optimal} linear device that attains the upper-bound performance of a nonlinear harvester~\cite{Halvorsen13}.

However, the concept of white noise is an idealisation that may not be valid in many practical situations. Random fluctuations acting on physical, chemical or biological systems actually have a finite correlation time. For example, in the classical Brownian process there is a timescale given by the typical collision time of the fluid molecules with the Brownian particle below which fluctuations cannot be considered uncorrelated. The existence of finite correlation times is even more important in complex fluids, where hydrodynamic fluctuations can be correlated over long time intervals. This is specially relevant for practical harvesting. Since an efficient harvester would require to have a response that peaks within the lower end of the frequency bandwidth, where most of the noise energy is concentrated ---and considering that there are physical limits to the mass or string constants that can be used to tune the resonant frequency of such harmonic oscillator--- it is unclear that the optimal harmonic harvester (see Ref.~\cite{Halvorsen13}) may be actually realizable in systems  where environmental fluctuations are characterized by colored noise.

After some early experimental and simulation studies~\cite{Vocca12,Nguyen11}, the power output of both a monostable~\cite{Daqaq10} and a bistable Duffing oscillator with a symmetric potential~\cite{Daqaq11} driven by Ornstein-Uhlenbeck noise was determined by approximate methods, and the exact analytical expressions for the net electrical power and efficiency of the conversion of the power supplied by exponentially correlated noise into electrical power was derived for a linear electromechanical oscillator employed as an energy harvester~\cite{Mendez13}.

Notwithstanding the recent advances in nonlinear vibration energy harvesters, some important issues have remained unaddressed so far. The mechanical part of these systems is usually modelled with a harmonic potential plus a nonlinear one that can be considered as an effective potential that accounts for the degrees of freedom not explicitly considered in the linear description. This issue becomes relevant at nanometric scales wherein the detailed structure of the mechanical resonator has to be taken explicitly into account. This is not only to construct the model, but also to assess the influence of these non-accounted for degrees of freedom in the dynamics. Some examples in this direction consist in the studies of nanowire resonators~\cite{Husain03} and nanoribbons designed for vibrational energy harvesting processes~\cite{LopezSuarez15}. 

In this paper we propose a new energy harvester system that is able to effectively extract energy from the low end part of the environmental (colored) noise spectrum, where most energy is available, while being linear, simple, and amenable to analytical treatment. Our model consists of a $N$ harmonic oscillator chain with one end in contact with the ambient reservoir, while the other end is attached to a transduction circuit. We show that this configuration is able to overcome the single harmonic oscillator efficiency, specially in the case of ambient noise with a finite correlation time. We find that the harmonic chain leads to a broad spectral response of the first oscillator ---the one in contact with the ambient--- that overlaps with that of the external noise. This leads to an optimal energy extraction from the latter, in sharp contrast with the narrow spectral response of single, linear-oscillator-based harvesters.
	
Furthermore, the harmonic lattice lends itself to analytical treatment. We have derived an analytical approximation that sheds light on the results of spectral analysis obtained by numerical simulations. Our analytical results for the harmonic chain help explaining why our proposed model outperforms the single oscillator case for the considered parameters, both in delivered power as well as in efficiency.

The rest of the paper is organized as follows: in Sec. II we present the model as well as our methodology. Numerical as well as analytical results are reported in Sec. III. Finally, in Sec. IV we discuss the results so far obtained and propose ways to continue this line of research.

\section{model and methodology\label{sec:Model}}
\subsection{Single oscillator model}

An energy harvester is a device that converts the power supplied by external noise into electrical energy. This process begins with the damped oscillator being driven by the external noise. Kinetic energy is then converted via a piezoelectric transducer mechanism into electrical energy that is then stored in a capacitor. We will begin reviewing the original implementation~\cite{Mendez13}, that from now on will be termed single-oscillator case. The mechanical part of the device is described by the dynamical equation for the momentum of the stochastically driven damped oscillator of mass $m$, which reads as
\be
\dot{p} + \gamma\dot{q} + {\cal F_{\mathrm{tran}}}(q,V) + kq = \xi(t),
\label{uno}
\ee
where $q$ is the displacement from the equilibrium position and $p$ is the momentum, with the dot standing for temporal derivative. In this equation $k$ is the harmonic constant, $\gamma$ is the linear damping coefficient, $\xi(t)$ is the random driving force, and ${\cal F}_{\mathrm{tran}}(q,V)$ is the transducer force due to the motion-to-electricity conversion mechanism, which depends on the geometry of the transducer and on how the circuit that implements the energy conversion cycle operates. It opposes to the motion, just as the friction force, and has its origin in the energy loss that occurs when kinetic energy is converted into electric energy. The simplest expression for this function is ${\cal F}_{\mathrm{tran}}(q,V)=k_v V$, where $k_v>0$ is a piezoelectric parameter and $V(t)$ is the electric voltage of the circuit. The dynamical equation for the voltage has the form $\dot V={\cal F}(p,V)-V/\tau_p$, where $\tau_p=R_{_L}{\mathcal C}$ is the time associated to the charging process of the piezoelectric element, which is larger than any other characteristic time of the system, ${\mathcal C}$ is the capacitance of the piezoelectric component, $R_{_L}$ is the load resistance, and ${\cal F}(p,V)$ is the connecting function with the oscillator. In the following the latter will be taken as ${\cal F}(p,V)=k_c p$, where $k_c$ is the coupling constant of the piezoelectric sample.

In this work we are considering a Ornstein-Uhlenbeck (OU) random force, with mean $\langle\xi\rangle=0$ and correlation $\langle\xi(t)\xi(t^{\prime})\rangle=\sigma^2\exp(-|t-t^{\prime}|/\tau_c)$, where $\sigma$ is the amplitude and $\tau_c$ is the correlation time. The limit $\tau_c\rightarrow0$ and $\sigma^2\rightarrow\infty$, with $D=\sigma^2\tau_c$ constant, corresponds to the white noise limit~\cite{Hanngi95}. In order to obtain a closed system of equations, it is a standard procedure to employ the equation $\dot\xi=-\xi/\tau_c+{\zeta}(t)/\tau_c$, where $\zeta(t)$ is a Gaussian white noise with zero mean and correlation $\langle\zeta(t)\zeta(t^{\prime})\rangle=2\sigma^2\tau_c\delta(t-t^{\prime})$. Therefore the complete set of equations reads as
\begin{subequations}
\begin{align}
 \dot q & = \frac{p}{m} \label{dos_a} \\
 \dot p & = -kq + \xi - \frac{\gamma}{m}p - k_v V \label{dos_b} \\
 \dot V & = \frac{k_c}{m}p - \frac{1}{\tau_p}V \label{dos_c} \\
 \dot\xi & = -\frac{\xi}{\tau_c} + \frac{\zeta}{\tau_c} \label{dos_d}.
\end{align}
\end{subequations}

Since the total mechanical energy of the oscillator is $E=p^2/2m+U(q)$, where $U(q)=kq^2/2$ is the harmonic potential, the corresponding instantaneous power is given by its temporal derivative, {\it i.e.} $\dot E=\dot q[\dot p+U^{\prime}(q)]$. If in this last expression we substitute Eq.~(\ref{dos_b}) and take the statistical average we obtain
\be
\langle\dot E\rangle=\langle\dot q\xi\rangle - \gamma\langle\dot{q}^2\rangle - k_v\langle\dot q V\rangle,
\ee
where $\langle\cdots\rangle$ implies both a time-average during the observation interval and an ensemble average over noise realizations. $\langle\dot q\xi\rangle$ is the power delivered by the noise, $\gamma\langle\dot{q}^2\rangle$ is the power dissipated by friction, and $k_v\langle\dot q V\rangle$ is the power transferred from the oscillator to the transducer. Next, if in $dV^2/dt=2V\dot V$ we substitute Eq.~(\ref{dos_c}), take the statistical average, and consider that the system is in the stationary regime ---thus $d\langle V^2\rangle/dt=0$---, we obtain the relation $\langle V^2\rangle=k_c\tau_p\langle \dot q V\rangle$. The transducer's efficiency in converting mechanical to electrical power is given by
\be
\eta_{\mathrm{me}}=\frac{\langle V^2\rangle/R_{_L}}{k_v\langle\dot{q}V\rangle}=\frac{k_c\tau_p\langle \dot q V\rangle/R_{_L}}{k_v\langle\dot{q}V\rangle}=\frac{k_c {\mathcal C}}{k_v}\le1.
\ee
Since the transduction mechanism is not the object of the present study we can consider, without loss of generality, that $\eta_{\mathrm{me}}=1$, which implies that $k_c=k_v={\mathcal C}=1$, being values that we will henceforth employ. The conversion efficiency of the power delivered by the noise to power transferred from the oscillator to the transducer is $\eta_{\mathrm{rm}}=k_v\langle\dot{q}V\rangle/\langle\dot{q}\xi\rangle$; thus the overall conversion efficiency of power delivered by the noise to net electrical power can be written as
\be
\eta=\eta_{\mathrm{me}}\eta_{\mathrm{rm}}=\frac{\langle V^2\rangle/R_{_L}}{\langle\dot{q}\xi\rangle}.
\ee

\subsection{Proposed model: Harmonic oscillator chain}

\begin{figure}\centering
\includegraphics[width=0.75\linewidth,angle=0.0]{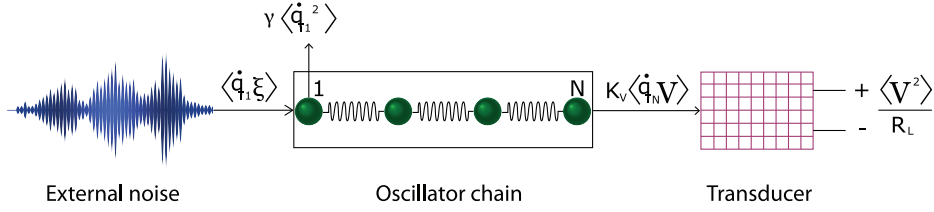} 
\caption{
	Sketch of an energy harvester based on a chain of harmonic oscillators with one end in contact with the ambient noise and the other attached to the electrical transducer circuit.}
\label{fig:1}
\end{figure}

As a mechanical resonator we now consider a one-dimensional chain of $N$ nearest-neighbor harmonic oscillators, as sketched in Fig.~\ref{fig:1}. Within this scheme the first oscillator in the lattice $q_{_1}$ is directly in contact with the stochastic signal $\xi(t)$ and the last one, $q_{_N}$, is connected to the transducer. Thus the equations of motion can be written as
\begin{subequations}
\begin{align}
 \dot q_i & = \frac{p_i}{m_i} \label{dosx_a} \\
 \dot p_i & = F_i + \delta_{i1}\Bigl(\xi - \frac{\gamma}{m_i}p_i\Bigr) - \delta_{iN}k_v V  \label{dosx_b} \\
 \dot V & = \frac{k_c}{m_{_N}}p_{_{N}} - \frac{1}{\tau_p} V \label{dosx_c} \\
 \dot\xi & = -\frac{\xi}{\tau_c} + \frac{\zeta}{\tau_c} \label{dosx_d},
\end{align}
\end{subequations}
where $F_i=-F(q_{i+1}-q_i)+F(q_i-q_{i-1})$ and $F(x)=-kx$ being the force on the $i$th oscillator due to the nearest-neighbor interaction within the lattice.

The mechanical energy is defined through the expectation value of the lattice Hamiltonian, that is,
\be
\langle E \rangle =\Biggl\langle\sum_{i=1}^N\frac{m_i}{2}\dot{q}_i^2+\sum_{i=1}^{N-1}\frac{k}{2}(q_{i+1}-q_i)^2\Biggr\rangle.
\ee
It can be immediately shown, employing Eq.~(\ref{dosx_b}), that
\be
\langle\dot{E}\rangle=\langle\dot{q_{_1}}\xi\rangle - \gamma\langle\dot{q}_{_1}^2\rangle - k_v\langle\dot{q}_{_N}V\rangle, \label{tres}
\ee
where $\langle\dot{q_{_1}}\xi\rangle$ is the power delivered by the noise, $\gamma\langle\dot{q_{_1}}^2\rangle$ is the power dissipated by friction with the first oscillator in the lattice, and $k_v\langle\dot{q}_{_N}V\rangle$ is the power transferred from the $N$th oscillator to the transducer. Since we have been able to maintain the analogy with the original single-oscillator case, we can infer that the total efficiency of the conversion process from power delivered by the external noise to final net electrical power can be defined as the quotient of both powers, and thus
\be
\eta=\frac{\langle V^2 \rangle/R_{_L}}{\langle\dot{q}_{_1}\xi\rangle},
\label{eff}
\ee
which is an expression completely analogous to the one previoulsy employed in the literature~\cite{Mendez13}. 

\section{Results\label{sec:SA}}

\subsection{Numerical simulations}

The simulations are performed by solving numerically the Langevin equations~(\ref{dos_a}-\ref{dos_d}) and (\ref{dosx_a}-\ref{dosx_d}) by using the so-called Heun algorithm; trajectories are computed over an interval of $4096$ time units after a transient of $10^3$ starting from a set of initial conditions given by $\{q(0)=p(0)=V(0)(\equiv V_{_0})=0\}$. An ensemble average over $10^3$ independent realizations has been performed for the chosen parameter set. 

\begin{figure}\centering
\includegraphics[width=0.75\linewidth,angle=-90.0]{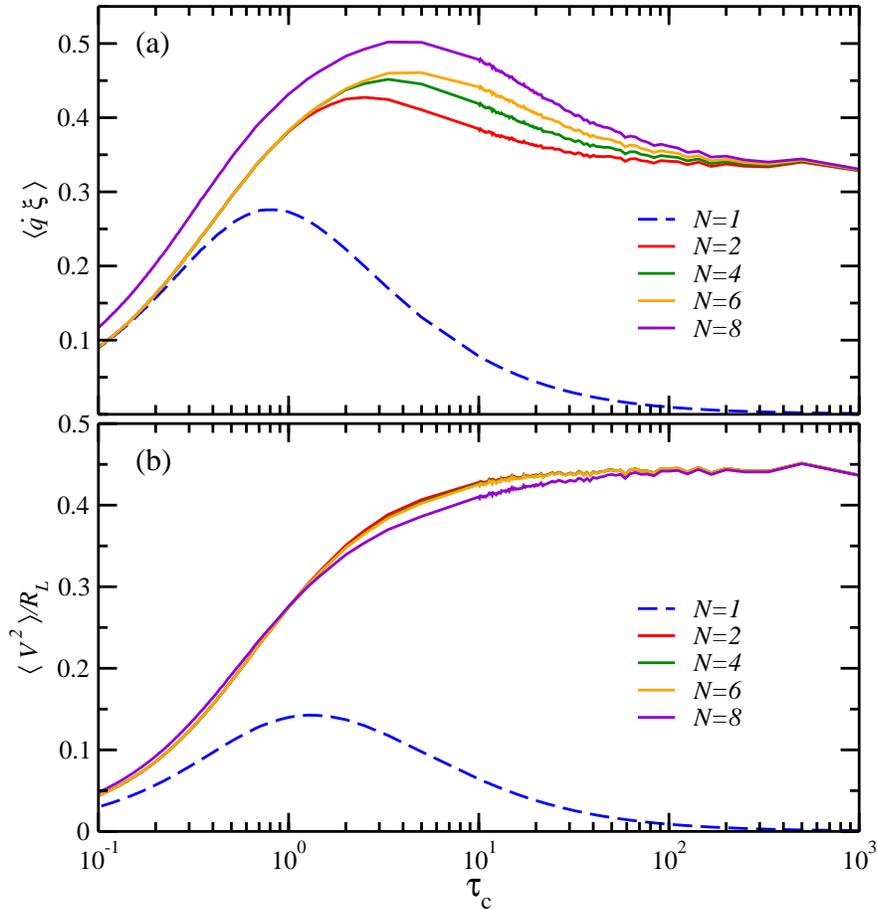} 
\caption{
	(a) Correlation $\langle\dot{q}\xi\rangle$ vs $\ct$ for an oscillator harvesting system with with $N=1$ (blue dashed line), $2$ (red), $4$ (green), $6$ (orange), and $8$ (violet). (b) Correlation $\langle V^2\rangle/R_{_L}$ vs $\ct$. In both panels the parameters are $\sigma^2=\gamma=k=k_c=k_v={\mathcal C}=1$, $V_{_0}=0$, $m_i=1\,\,\forall\,i$, and $\tau_p=R_{_L}=2$ in all considered instances.}
\label{fig:2}
\end{figure}

In Fig.~\ref{fig:2}(a) we present the behavior of the power delivered by the external noise as a function of the correlation time. For the single-oscillator case it is clear that external energy can be significantly harvested only arround a definite value of $\ct\approx1$, and in a rapidly decreasing rate in both small and large correlation time limits. But in contrast, for the oscillator chain the energy harvested only drops in the white-noise limit, {\it i.e.} for very short correlation times. In the opposite limit the delivered power is markedly higher than that from the single-oscillator harvester for all system sizes and $\ct$ values considered. On the other hand, the net electrical power $\langle V^2\rangle/R_{_L}$ depicted in Fig.~\ref{fig:2}(b) presents, for the single oscillator instance, a very similar behavior as its corresponding delivered power: it has a maximum at a $\ct\approx1$ value and is sub-optimal in the entire $\ct$ value range. But for $N>1$ sizes the net electrical power only decreases in the white-noise limit. By contrast, it seems to become independent of both $\ct$ and $N$ in the colored noise limit, {\it i.e.} large $\ct$ values. And again, for all correlation time values considered the net electrical power is much higher than that delivered by the single oscillator case. Therefore the oscillator chain, even for a small value of $N=2$, outperforms the single oscillator energy harvester both in delivering power to the system as well as in rendering net electrical power.

\begin{figure}\centering
\includegraphics[width=0.75\linewidth,angle=0.0]{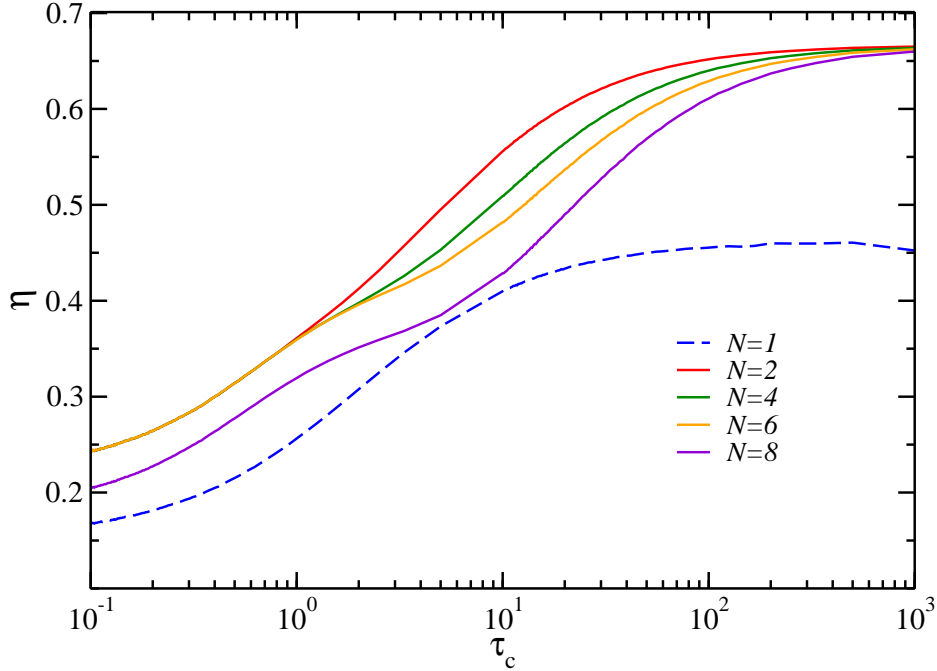}
\caption{
	Efficiency $\eta$ vs $\ct$ for the system sizes, parameter values, and color coding as in Fig.~\ref{fig:2}.}  
\label{fig:3}
\end{figure}

The efficiency dependence on $\ct$ is displayed in Fig.~\ref{fig:3}. For all the considered instances the highest efficiency figure is achieved in the large $\ct$ limit. However, the systems with $N>1$ outperform the single-oscillator one in the whole studied value range, being particularly efficient in the range $\ct>10^2$, wherein a decreasing $N$ dependence can be noticed. Furthermore, even if the single-oscillator $N=1$ harvester has also an almost $\ct$-independent efficiency in that same value range, the corresponding net electrical power, see Fig.~\ref{fig:2}(b), is insignificant, something that does not happen when larger system sizes are considered.

\begin{figure}\centering
\includegraphics[width=0.75\linewidth,angle=0.0]{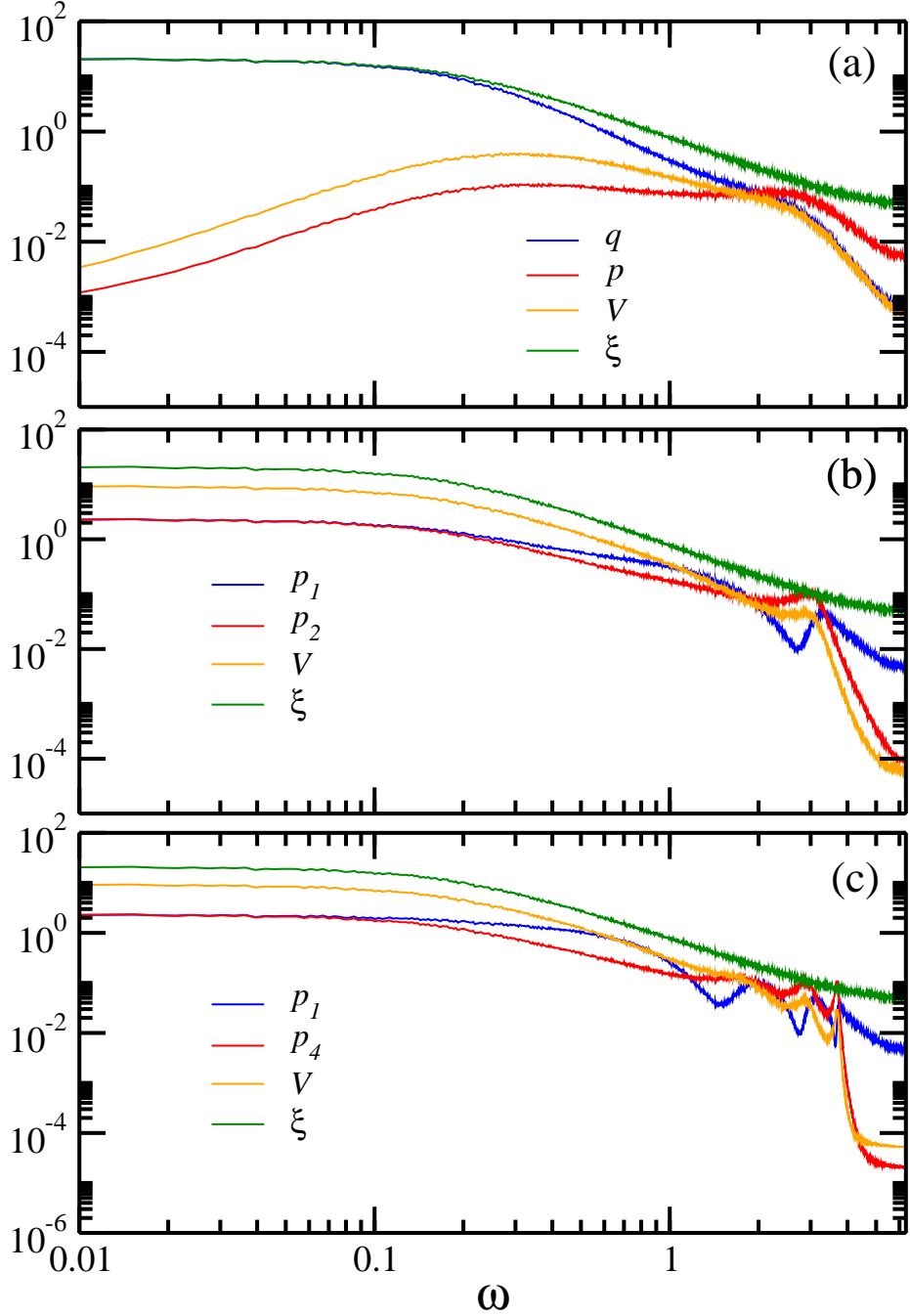}
\caption{
	(a) Power spectra $\langle|\hat{X}(\omega)|^2\rangle$ for the single oscillator system variables $\{X\}$: displacement (blue), momentum (red), voltage (orange), and the OU noise (green). (b) Power spectra for an oscillator lattice with $N=2$ corresponding to $p_{_1}$ (blue), $p_{_N}$ (red), voltage (orange), and the OU noise (green). (c) Same as in (b) but for $N=4$. In all instances $\ct=10$. Same parameter values as in Fig.~\ref{fig:2}.}
\label{fig:4}
\end{figure}

In order to elucidate the origin of the poor performance of the single-oscillator energy harvester we carry out a spectral analysis of all the dynamical variables involved. Thus in Fig.~\ref{fig:4}(a) we present the corresponding power spectra. Taking into account the average power balance in Eq.~(\ref{tres}) it is clear that the relevant correlations correspond to those of $\dot q$ with the external noise $\xi$ and the voltage $V$, since they are related to the power delivered to the noise into the transducer and to the net electrical power, respectively. Now, in the low-frequency limit the velocity has a power-law dependence $\sim\omega^{2}$ that rapidly decouples its behavior to that of $\xi$, thus decreasing the amount of delivered power and reducing the performance of the device. The velocity is coupled to the voltage, but because the former is decoupled from the noise at low frequencies, the latter shares the aforementioned decay; thus a low net electrical power is obtained in this case. However, for $N=2$ both $\dot q_{_2}$ and $V$ are now coupled to the behavior of the external noise at low frequencies, which assures a significant correlation in that frequency regime. This spectral behavior explains the non-decaying correlations depicted in Fig.~\ref{fig:2} and the sizable efficiency in Fig.~\ref{fig:3} for $\ct>10^2$ values. An additional advantage is that this result is robust for larger system sizes, since the spectra in Fig.~\ref{fig:4}(c) for $N=4$ remain mostly unchanged.

\subsection{Analytical results}

Next, to gain a deeper understanding of the low-frequency behavior depicted in Fig.~\ref{fig:4} we perform an analytical study of the harvester in the stationary regime. By taking a sufficiently long transient time we can neglect the contribution from the lattice initial state given that normal modes decay after some time. This allows us to solve Eqs.~(\ref{dos_a}-\ref{dos_d}) via Fourier transform. Let us denote $\hat{X}(\omega) \equiv \int dt X(t) \exp(-\mathrm{i}\omega t)$; then Eqs.~(\ref{dos_b}-\ref{dos_c}) can be written as
\begin{subequations}
\begin{align}
-m\omega^2 \hat q(\omega) + \mathrm{i}\gamma\omega\hat q(\omega) + k_v\hat{V}(\omega) + k\hat q(\omega) & = \hat\xi(\omega) \label{xx_a} \\
 \mathrm{i}\omega\hat {V}(\omega) = \mathrm{i}k_c\omega\hat q(\omega) - \frac{\hat{V}(\omega)}{\tau_p}. \label{xx_b}
\end{align}
\end{subequations}
After substituting Eq.~(\ref{xx_b}) in (\ref{xx_a}) we can calculate the power spectrum of $\hat q(\omega)$ as
\be
|\hat{q}(\omega|^2=\Big(\frac{1}{A_0(\omega)^2+B_0(\omega)^2}\Big)|\hat\xi(\omega)|^2 \label{yy},
\ee
where
\be
A_0(\omega)=(k-m\omega^2) + \frac{k_vk_c(\tau_p\omega)^2}{1+(\tau_p\omega)^2}
\ee
and
\be
B_0(\omega)=\gamma\omega + \frac{k_vk_c\tau_p\omega}{1+(\tau_p\omega)^2}.
\ee
From Eq.~(\ref{yy}) it is clear that the behavior of the displacement and the external noise are closely correlated for any frequency value, as can indeed be corroborated from the data reported in Fig.~\ref{fig:4}(a). In particular, in the low-frequency regime
\be
\frac{1}{A_0(\omega)^2+B_0(\omega)^2}\sim\frac{1}{k^2}\Big(1-C\Big[\frac{\tau_p\omega}{k}\Big]^2\Big),
\ee
with the constant $C=2k(k_vk_c-m/\tau_p^2)+(\gamma/\tau_p+k_vk_c)^2$. Then, from this last expression it is clear that, in the $\omega\rightarrow0$ limit, $|\hat q(\omega)|^2\sim k^{-2}|\hat\xi(\omega)|^2$. This is precisely the coupling of the displacement and external noise that can be appreciated in Fig.~\ref{fig:4}(a). As for the momentum we have $|\hat{p}(\omega)|^2 \propto|\hat{\dot q}(\omega)|^2\sim (\omega /k)^2|\hat\xi(\omega)|^2$, which is again the behavior displayed in the aforementioned figure. From the close coupling between the displacement and voltage inferred from Eq.~(\ref{xx_b}) it is immediate to deduce that the output voltage will experience a drastic drop in the low-frequency region, which prevents the harvester to adequately perform in the long correlation time limit $\ct\gg1$.

For the analytical treatment of the chain we employ the methodology recently developed in Ref.~\cite{Weiderpass20} based on the finite version of the so-called $Z$-transform, which allows to obtain closed expressions for $\hat q_{_1(\omega)}$ and $\hat q_{_N}(\omega)$ in terms of $\hat\xi(\omega)$ that read as
\begin{subequations}
\begin{align}
\hat q_{_1}(\omega) & = \frac{B(\omega)\hat q_{_N}(\omega)-\hat\xi(\omega)}{\omega D(\omega)}\label{yy_a}\\
	\hat q_{_N}(\omega) & = E(\omega)\hat{q_{_1}(\omega)} - \Big(\frac{N-1}{k}\Big)\hat\xi(\omega), \label{yy_b}
\end{align}
\end{subequations}
where
\begin{subequations}
\begin{align}
B(\omega)&=\frac{k_vk_c\omega\tau_p}{\tau_p\omega-\mathrm{i}},\\
D(\omega)&=m\omega N + \mathrm{i}\gamma,\\
E(\omega)&=1-\mathrm{i}\gamma\frac{\omega}{k}(N-1).
\end{align}
\end{subequations}
These are the corresponding expressions of Ref.~\cite{Weiderpass20} in the low-frequency limit and particularized to our system. 

Therefore, after substituting Eq.~(\ref{yy_b}) into Eq.~(\ref{yy_a}), now we can express $\hat q_{_1}(\omega)$ in terms of the power spectrum of the external noise as
\be
|\hat q_{_1}(\omega)|^2=\Big|\frac{1+B(\omega)(N-1)/k}{\omega D(\omega)-B(\omega)E(\omega)}\Big|^2|\hat\xi(\omega)|^2.
\ee
After some algebra, it can be shown that the order of magnitude of the prefactor in the last equation has the form
\be
|\hat q_{_1}(\omega)|^2\sim\Big(\frac{1+{\mathcal O}(\omega^2)}{{\mathcal O}(\omega^2)}\Big)|\hat\xi(\omega)|^2.
\ee

On the other hand, for the last oscillator, from Eq.~(\ref{yy_b}) we obtain
\be
|\hat q_{_N}(\omega)|^2  = |E|^2|\hat q_{_1}|^2+G^2|\hat\xi|^2 -
 2 G \, \operatorname{\mathbb{R}e}(E\hat q_{_1}\hat\xi^{*}).
\ee

And again, after some lengthy algebra we obtain the expansion for $\omega \to 0$
\be
|\hat q_{_N}(\omega)|^2\sim\Big(\frac{1+{\mathcal O}(\omega^2)}{{\mathcal O}(\omega^2)}+\mathrm{const.}\Big) \, |\hat\xi(\omega)|^2,
\ee
and the corresponding momenta of both boundary oscillators, at lowest order in $\omega$, depend on the frequency as
\be
|\hat{p}_{_{1,N}}(\omega)|^2\sim\Big(1+{\mathcal O}(\omega^2)\Big) \, |\hat\xi(\omega)|^2,
\ee
which renders $|\hat{p}_{_{1,N}}(\omega)|^2 \propto |\hat\xi(\omega)|^2$ as $\omega \to 0$, a behavior clearly corroborated by the simulation results in Figs.~\ref{fig:4}(b) and (c). 

Finally, since from Eq.~(\ref{xx_b}) we have that $\hat V(\omega)=[k_c\omega\tau_p/(\omega\tau_p-\mathrm{i})] \, \hat q_{_N}(\omega)$, employing Eq.~(\ref{yy_b}) we obtain the approximation
\be
|\hat V(\omega)|^2\sim(k_c\tau_p)^2 \Big(1+\mathcal{O}(\omega^2)\Big) \, |\hat\xi(\omega)|^2,
\label{chain_elec_pot}
\ee
for $\omega \to 0$, which is in excellent  agreement with the simulation results in Figs.~\ref{fig:4}(b) and (c).

In summary, our analytical results show that, in the low-frequency limit, the single harmonic oscillator harvester is able to extract energy from the noise generating an electric potential with a power spectrum that decays as $|\hat{V}(\omega)|^2 \sim \omega^2|\hat\xi(\omega)|^2 $, leading to less energy being harvested at lower frequencies. In contrast, for the $N$-oscillator chain, Eq.~(\ref{chain_elec_pot}), one has a flat spectrum $|\hat{V}(\omega)|^2 \sim |\hat\xi(\omega)|^2$. Since the total power transferred from the oscillators to the transducer is given by $\langle V^2 \rangle = \int d\omega |\hat{V}(\omega)|^2$, these results explain the much better performance of the harvester based on a chain as compared with the single oscillator, as shown by the numerical results in Fig.~\ref{fig:2} (b) and Fig.~\ref{fig:3}.

\section{discussion and conclusions\label{sec:Disc}}

Our results for the linear oscillator lattice electromechanical energy harvester interacting with an external finite-bandwidth ambient noise clearly show that its performance is enhanced, both in the net electrical power delivered and in its efficiency, compared with the single oscillator instance for any finite value of the noise correlation time $\ct$. For sufficiently large values of $\ct$, both net electrical power and efficiency become constant and take large values, in sharp contrast to the single oscillator energy harvester, where the combined goals of both maximum power and efficiency cannot be attained simultaneously. By means of spectral analysis we have elucidated the origin of the poor performance of the single oscillator energy harvester: a power-law frequency dependency $\sim\omega^2$ of the velocity power spectrum that renders its contribution negligible in the low-frequency limit. This, in turn, reduces significantly the power that can be harvested from the external noise. On the contrary, for the chain system resonance with the extra frequencies afforded by the additional degrees of freedom contributes to a non-decaying velocity/momentum power-spectrum in the low-frequency region, wherein most of the noise energy resides, thus rendering a consistent performance of the device for finite correlation time values. These numerical findings have been corroborated by an analytical approximation, with excellent agreement between both.

While most studies of energy harvesters typically consider uncorrelated environmental noise, the reality is that this limit is an idealisation to describe noises correlated over very short times. However, in many potential applications, like electromagnetic plasmas~\cite{Castellanos_2013}, non-Newtonian fluids~\cite{non-newtonian} or nanofluidics~\cite{nanofluidics}, the noise fluctuations may exhibit long correlation times. Therefore, energy harvesters that can take advantage of the low end frequency band without the need of fine tuning the device response frequency to the right bandwidth, are most welcome. In this respect, the chain of harmonic oscillators, with its flat response spectrum, can be a very effective, yet simple, way to harvest considerable amounts of energy.

As for possible experimental implementations we recall that the phonon mean-free path in graphene ($\sim$775 nm near room temperature~\cite{Balandin11}) is much longer than the sizes of various graphene nanostructures recently considered. Therefore, the intrinsic nonlinearity is insignificant and thus can be regarded as harmonic systems. Furthermore, since graphene has a very high thermal conductivity~\cite{Balandin11,Xu13}, its energy transport properties are quasi-ballistic, another property of harmonic systems. Besides graphene other materials with high thermal conductivity such as carbon nanotubes~\cite{Pop06,Donadio09} or carbyne~\cite{Wang15,Shi16} could be considered. At these nanoscopic scales it is known~\cite{Eichler11}, from studies of mechanical resonators based on carbon nanotubes~\cite{Husain03,Sazonova04} and graphene sheets~\cite{Bunch07,Chen09}, that damping strongly depends on the amplitude of motion and is better described by a nonlinear rather than the linear damping force used in the present study. Such nonlinearity leads to a broadening of the resonance frequency that most certainly will have a significant influence on the performance of the herein proposed energy harvester, which we plan to address in a future work.
 
\smallskip
\begin{acknowledgments}
M.~R.~B. gratefully acknowledges CONACyT, M\'exico for financial support during a sabbatical leave at IFCA/Universidad de Cantabria and to the latter for its warm hospitality. J.\ M.\ L. is partially supported by project No. FIS2016-74957-P from Agencia Estatal de Investigaci\'on (Spain) and FEDER (EU).
\end{acknowledgments}



\begin{thebibliography}{10}
	
	\bibitem{Paradiso05}
	J.~A. Paradiso and T. Starner, IEEE Pervasive Comput. {\bf 4},  18  (2005).
	
	\bibitem{Okamoto12}
	H. Okamoto, Y. Hamate, L. Xu, and H. Kuwano, Smart Materials and Structures
	{\bf 21},  065001  (2012).
	
	\bibitem{Kwon13}
	S.-D. Kwon, Y. Park, and K. Law, Smart Materials and Structures {\bf 22},
	055007  (2013).
	
	\bibitem{Anton07}
	S.~R. Anton and H.~A. Sodano, Smart Materials and Structures {\bf 16},  R1
	(2007).
	
	\bibitem{Renno09}
	J.~M. Renno, M.~F. Daqaq, and D.~J. Inman, Journal of Sound and Vibration {\bf
		320},  386   (2009).
	
	\bibitem{Kim11}
	H.~S. Kim, J.-H. Kim, and J. Kim, International Journal of Precision
	Engineering and Manufacturing {\bf 12},  1129  (2011).
	
	\bibitem{Zeng13}
	W. Zeng {\it et~al.}, Energy and Environmental Science {\bf 6},  2631  (2013).
	
	\bibitem{Beeby06}
	S.~P. Beeby, M.~J. Tudor, and N.~M. White, Measurement Science and Technology
	{\bf 17},  R175  (2006).
	
	\bibitem{Spreemann12}
	D. Spreemann and Y. Manoli, {\em Electromagnetic Vibration Energy Harvesting
		Devices: Architectures, Design, Modeling and Optimization} (Springer,
	Dordrecht, 2012).
	
	\bibitem{Burrow08}
	S.~G. Burrow, L.~R. Clare, A. Carrella, and D. Barton,  in {\em Proceedings of
		the SPIE smart structures/NDE conference}, edited by M. Ahmadian (SPIE, San
	Diego, California, USA, 2008).
	
	\bibitem{Wang08a}
	Z.~L. Wang, Sci. Am. {\bf 82},  298  (2008).
	
	\bibitem{Cottone09}
	F. Cottone, H. Vocca, and L. Gammaitoni, Phys. Rev. Lett. {\bf 102},  080601
	(2009).
	
	\bibitem{Gammaitoni09}
	L. Gammaitoni, I. Neri, and H. Vocca, Appl. Phys. Lett. {\bf 94},  164102
	(2009).
	
	\bibitem{Erturk09}
	A. Erturk, J. Hoffmann, and D.~J. Inman, Appl. Phys. Lett. {\bf 94},  254102
	(2009).
	
	\bibitem{Harne13}
	R. Harne and K. Wang, Smart Mater. Struct. {\bf 22},  023001  (2012).
	
	\bibitem{Hosseinloo16}
	A.~H. Hosseinloo and K. Turitsyn, Smart Mater. Struct. {\bf 25},  015010
	(2016).
	
	\bibitem{Hosseinloo15}
	A.~H. Hosseinloo and K. Turitsyn, Phys. Rev. Appl. {\bf 4},  064009  (2015).
	
	\bibitem{Daqaq10}
	M.~F. Daqaq, Journal of Sound and Vibration {\bf 329},  3621   (2010).
	
	\bibitem{Green12}
	P.~L. Green, K. Worden, K. Atallah, and N.~D. Sims, Journal of Sound and
	Vibration {\bf 331},  4504   (2012).
	
	\bibitem{Halvorsen13}
	E. Halvorsen, Phys. Rev. E {\bf 87},  042129  (2013); {\em Erratum}, Phys. Rev. E {\bf 88}, 039902 (2013).
	
	\bibitem{Vocca12}
	H. Vocca, I. Neri, F. Travasso, and L. Gammaitoni, Applied Energy {\bf 97},
	771   (2012).
	
	\bibitem{Nguyen11}
	S.~D. Nguyen and E. Halvorsen, Journal of Microelectromechanical Systems {\bf
		20},  1225  (2011).
	
	\bibitem{Daqaq11}
	M.~F. Daqaq, Journal of Sound and Vibration {\bf 330},  2554   (2011).
	
	\bibitem{Mendez13}
	V. M\'endez, D. Campos, and W. Horsthemke, Phys. Rev. E {\bf 88},  022124
	(2013); {\em Erratum}, Phys. Rev. E {\bf 91}, 029904 (2015).
	
	\bibitem{Husain03}
	A. Husain {\it et~al.}, Appl. Phys. Lett. {\bf 83},  1240  (2003).
	
	\bibitem{LopezSuarez15}
	M. L\'opez-Su\'arez, G. Abadal, L. Gammaitoni, and R. Rurali, Nano Ene. {\bf
		15},  329–334  (2015).
	
	\bibitem{Hanngi95}
	P. H\"anngi and P. Jung, {\em Advances in Chemical Physics, Volume LXXXIX}
	(John Wiley Sons, Inc., New York, 1995), p.\ 239.
	
	\bibitem{Weiderpass20}
	G.~A. Weiderpass, G.~M. Monteiro, and A.~O. Caldeira, arXiv.org:2002.05195v1
	(2020).
	
	\bibitem{Castellanos_2013}
	O. Castellanos, J.~M. L{\'{o}}pez, J.~M. Sent{\'{\i}}es, and E. Anabitarte, J.
	Stat. Mech. Theory Exp. {\bf 2013},  P04022  (2013).
	
	\bibitem{non-newtonian}
	T. Burghelea and V. Bertola, {\em Transport Phenomena in Complex Fluids}
	(Springer, Cham, 2019).
	
	\bibitem{nanofluidics}
	I. Carrillo-Berdugo {\it et~al.}, Sci. Rep. {\bf 9},  7595  (2019).
	
	\bibitem{Balandin11}
	A.~A. Balandin, Nat. Mater. {\bf 10},  569  (2011).
	
	\bibitem{Xu13}
	X. Xu {\it et~al.}, Nat. Commun. {\bf 5},  3689  (2014).
	
	\bibitem{Pop06}
	E. Pop {\it et~al.}, Nano Lett. {\bf 6},  96  (2006).
	
	\bibitem{Donadio09}
	D. Donadio and G. Galli, Phys. Rev. Lett. {\bf 99},  255502  (2009).
	
	\bibitem{Wang15}
	M. Wang and S. Li, Sci. Rep. {\bf 5},  18122  (2015).
	
	\bibitem{Shi16}
	L. Shi {\it et~al.}, Nat. Mater. {\bf 15},  634  (2016).
	
	\bibitem{Eichler11}
	A. Eichler {\it et~al.}, Nat. Nanotech.  339  (2011).
	
	\bibitem{Sazonova04}
	V. Sazonova {\it et~al.}, Nature {\bf 431},  284  (2004).
	
	\bibitem{Bunch07}
	J.~S. Bunch {\it et~al.}, Science {\bf 315},  490  (2007).
	
	\bibitem{Chen09}
	C. Chen {\it et~al.}, Nature Nanotech. {\bf 4},  861  (2009).
	
\end{thebibliography}

\section{AUTHOR CONTRIBUTIONS STATEMENT} All authors contributed equally in all aspects of this work.

\end{document}